# Hydrogen adsorption on Pd(133) surface

Agnieszka Tomaszewska, Zdzisław M. Stępień,
*Institute of Physics, Jan Długosz University*
*al. Armii Krajowej 13/15, 42-200 Częstochowa, Poland*
*e-mail address: a.tomaszewska@ajd.czest.pl*

In this study used is an approach based on measurements of the total energy distribution (TED) of field emitted electrons in order to examine the properties of Pd (133) from the aspect of both hydrogen adsorption and surface hydrides formation. The most favourable sites offered to a hydrogen atom to be adsorbed have been indicated and an attempt to describe the peaks of the enhancement factor R spectrum to the specific adsorption sites has also been made.

## I. Introduction

Hydrogen adsorption on d transition metal surfaces, owing to its particular importance in catalysis, has attracted considerable attention for many years and a lot of works have been made in order to investigate phenomena that accompany the interaction of interest.
The H/Pd adsorption system is considered as a very convenient object for examination of technologically important processes taking place on the surface and a lot of efforts to study palladium surfaces interacting with hydrogen have been undertaken. Most of those works, both experimental and theoretical, were addressed to well-defined low index crystallographic planes [1-11]. It was determined that hydrogen adsorption on Pd has a dissociative character [1] and at temperatures 115-140 K and pressure as low as $10^{-4}$ Pa hydrogen diffusion into subsurface positions takes place [2,12].

Although it is well known that the presence of surface defects such as steps significantly influences the adsorption properties of the system [13], very little attention is paid to the hydrogen adsorption on defect-rich palladium surfaces. Very few exceptions concern Pd (210) [14-15] as well as Pd (311) [16].

In this paper we focus on hydrogen interaction with the Pd (133) surface whose adsorption properties, due to the existence of a wide variety of potential adsorption sites with different coordination numbers, are expected to be unique. A model of the surface under study is shown in Fig.1. It seems to be a stepped one which consists of terraces of close packed (111) plane.
We shall discuss here the results of the total energy distribution (TED) investigation of electrons field emitted from the palladium field-emitter tip surface exposed to hydrogen which were partially presented in our previous study [17]. The origin of TED peaks is discussed in reference to the structure of (133) surface. In particular, an attempt to identify the energetically most advantageous adsorption sites available to a hydrogen atom is undertaken. The possibility of surface hydride formation is also considered in the context of the observed alterations in both TED picture and field-emission microscope (FEM) image.

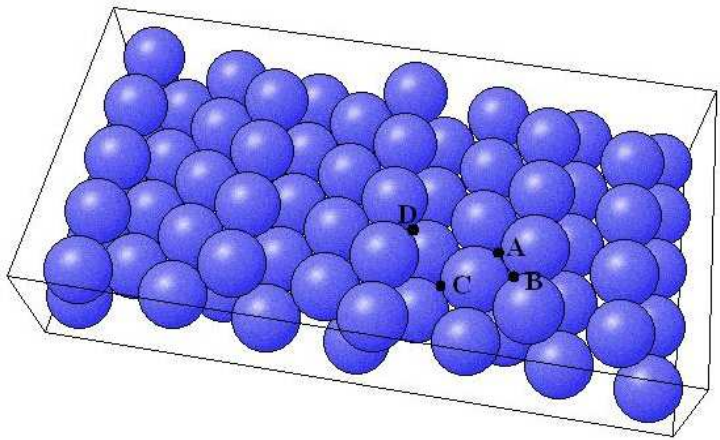

Fig.1 Top view of the Pd (133) surface. The most probable adsorption sites are marked by A, B, C, D.

## II.   Experimental

The emitter tip was prepared from a 0.1 mm palladium wire of 99.99 % purity by means of AC (~ 5 V) electropolishing in 38 % solution of HCl in glycerine [18]. Emitter was spot welded to a molybdenum 0.2-mm-wire loop. The loop was mounted on the cold finger so that the emitter was kept at liquid nitrogen temperature during the course of imaging. The cooled emitter tip assemble was put into the FEM microscope which was equipped with a standard ultrahigh vacuum system as well as low-pressure measurement facilities.

Prior to measurements, the emitter surface was cleaned through pre-annealing to a temperature of 1000 K. During the course of measurements the palladium surface was exposed to $H_2$ doses at 78 K. After the background pressure $10^{-10}$ hPa was reached, the emitter surface was exposed to hydrogen, but as before each successive dose of hydrogen was introduced into the vacuum chamber the emitter was heated up to 1000 K and the system was pumped out.

The FEM patterns were registered by means of digital photo-camera. TED measurements were recorded using a retarding potential analyzer being a slight modification of the Plummer and Young one [19]. The analyze's resolution, estimated by the correlation theory [20], was about 50 meV. However, it should be underlined that our attention is paid mainly to the qualitative description of hydrogen adsorption on Pd (133) surface. Hence, all the quantitative data presented here are of tentative meaning.

## III.   Results and discussion

Fig.2 shows an FEM pattern of the (111)-oriented clean palladium surface as well as a series of images taken after increasing hydrogen exposures of the surface at $T$= 78 K. Prior to the exposure of the palladium surface to hydrogen, a strong electron emission from the regions around the {110}-type planes is observed (Fig.2a). For a hydrogen exposure of 10 L strong deformation of the regions in question is noticeable (Fig.2b). This deformation manifests itself by the formation of individual areas of oblong shape. Increasing the applied exposure to 240 L leads to the appearance of a large number of planes which are placed on a perimeter of triangle in the centre of which the (111) plane is positioned (Fig.2c). In the final picture (Fig.2d) which depicts the situation at the exposure of 600 L, the (111) plane is surrounded by the ring composed of separate particles corresponding to the planes that contribute greatly to the total emission current.

a)

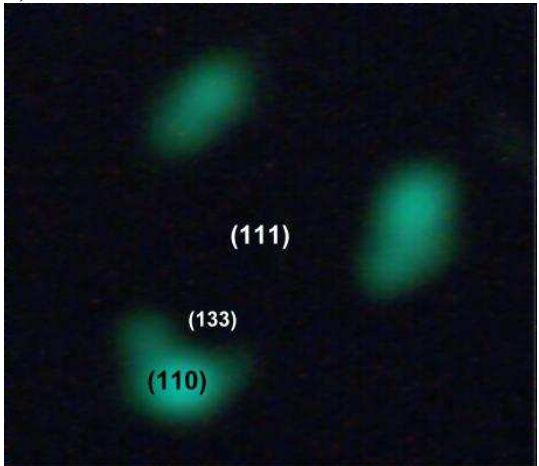

b)

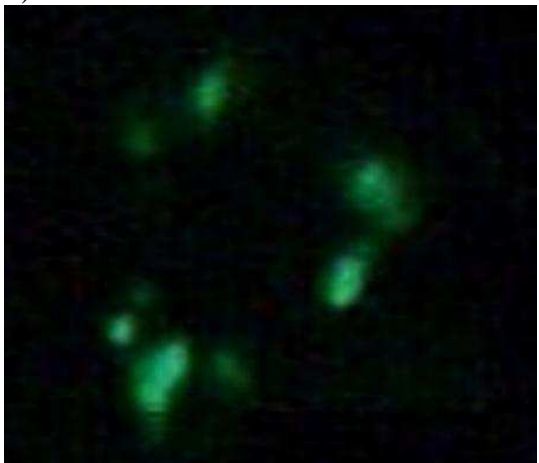

c)

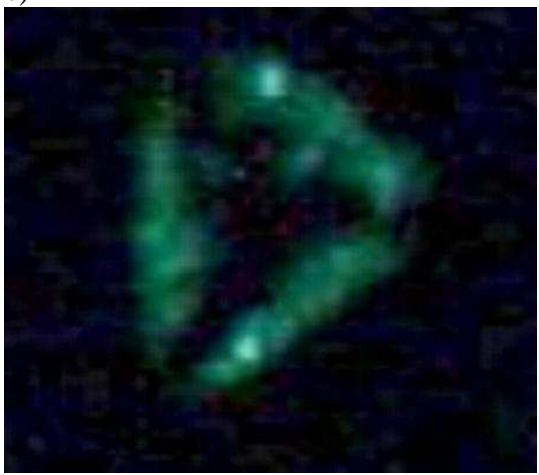

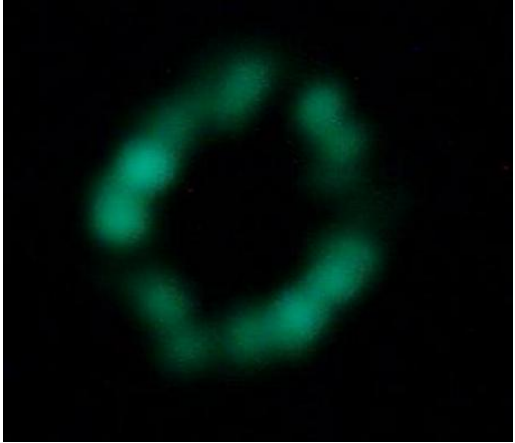

Fig.2 a) FEM images of the clean (111)-oriented clean palladium surface taken after heating to 1000 K. The background pressure was of the order of $10^{-10}$ hPa. The set of FEM images taken after Pd surface exposure to hydrogen. The applied exposures are: 10L (b), 240 L (c), 600 L (d).

The observed alterations in FEM picture of the hydrogen exposed palladium tip are due to the appearance of the adsorbate-induced structural changes on the surface. However, on the basis of only these patterns we are not able to determine whether the changes of interest were caused exclusively by hydrogen chemisorption on the surface or by hydrogen diffusion into the subsurface interstitial positions, moreover, palladium hydrides formation must be taken into account in this case.

In order to give a more detailed specification of the hydrogen adsorption on the palladium surface, the measurements of the total energy distribution (TED) of field emitted electrons were carried out. The Pd (133) plane, whose structure is described in Introduction, was chosen for the local investigation. The choice of the plane of interest was not accidental. Namely, a field ion microscopy (FIM) pattern taken after Pd (133) surface exposure to hydrogen (Fig.3) reveals the existence of stripes whose origin was explained by surface palladium hydrides formation [21].

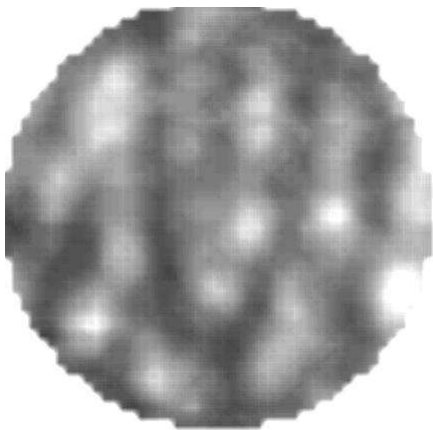

Fig.3 FIM pattern taken after Pd(133) exposure to hydrogen, obtained with krypton image gas ($p=10^{-3}$ hPa, U=25.5 kV) [21].

Hence, by employing the TED measurements of electrons emitted from the plane under study a great opportunity to examine changes in electronic structure of the substrate, induced by the process of surface hydrides formation, is provided.

Here, we focus on the interpretation of our findings from the aspect of surface hydrides formation. However, a brief presentation of the obtained results must be reminded. As for the clean surface, TED spectrum of emitted electrons (Fig.4a) does not reveal the presence of any resonant states. Hence, the effects connected with adsorption of residual gas cannot be taken into account in further interpretation. The energy of the Fermi level ($E_F$) was estimated to be about 4.25 eV.

a)

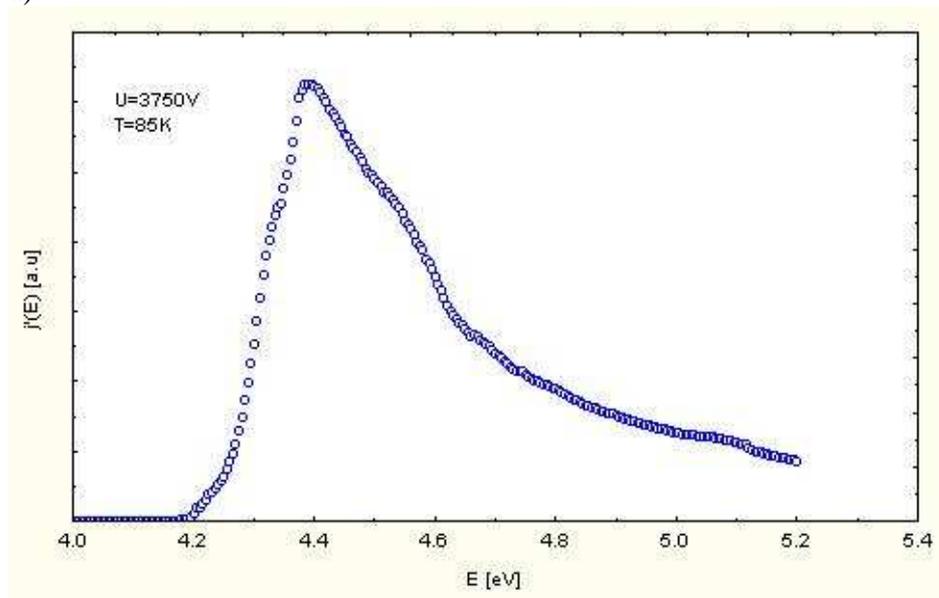

b)

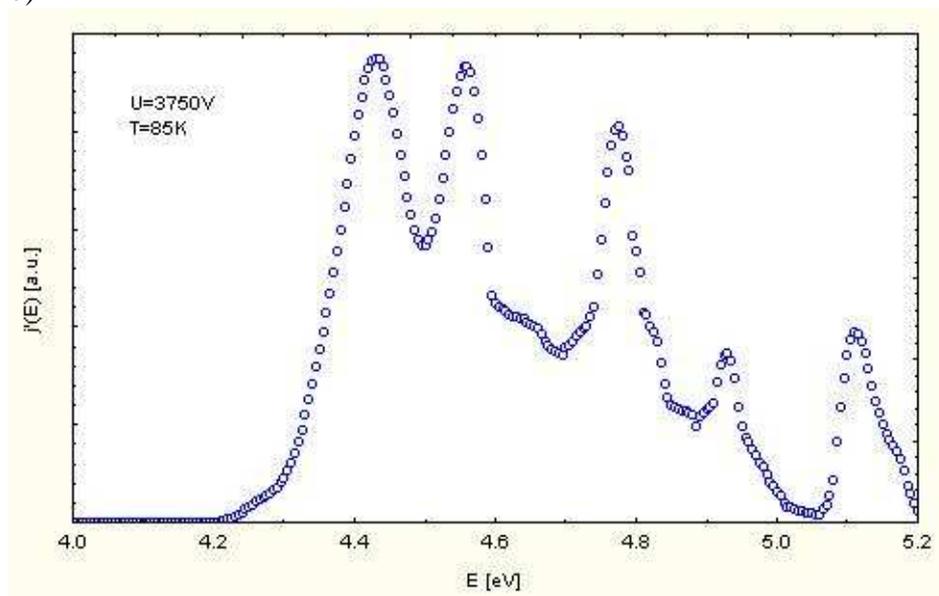

c)

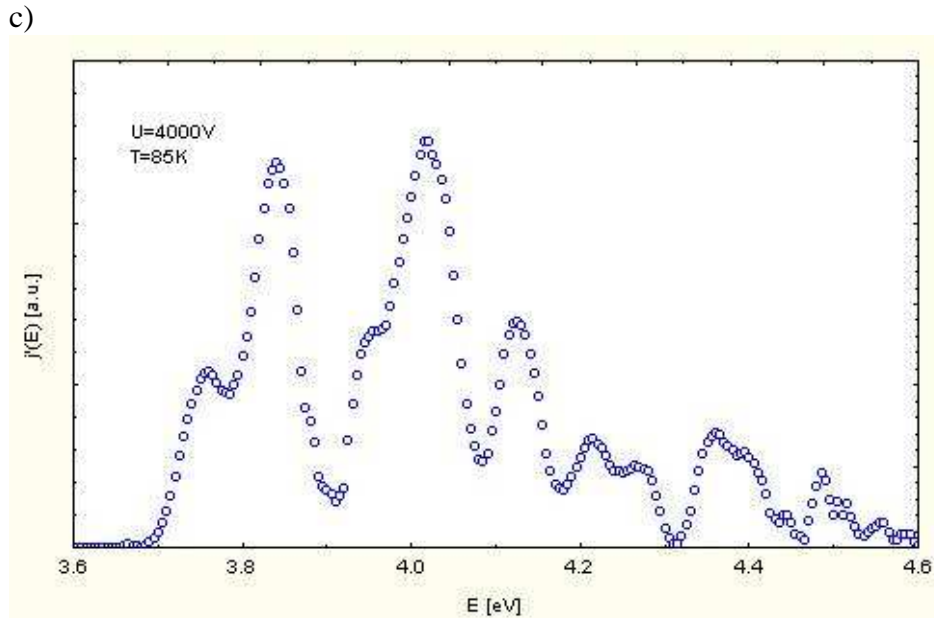

Fig.4 TED spectra of electrons field emitted from the clean (a) and hydrogen-adsorbed (b, c) Pd(133) surface. The hydrogen exposures are 240 L (b) and 600 L [17].

Adsorption of hydrogen at low temperature induces alterations in TED picture whose character depends on the amount of applied gas dose. After a hydrogen dose of 240 L was introduced, the appearance of four resonant states located at 0.32, 0.55, 0.70 and 0.88 below the $E_F$ was registered (Fig.4b). Within the commonly acceptable chemisorption model [22], the effects like this have to be explained by the appearance of four different chemisorption states. Such an explanation, with reference to the geometrical structure of the substrate on which the adsorption processes take place, seems to be reasonable. Namely, as it was mentioned in Introduction, the Pd (133) surface provides a large number of adsorption sites whose coordination varies from one to three. However, according to the generally approved rules, independently of the surface geometry, the sites with the highest coordination are most likely to be occupied [1]. Hence, in our case, only the sites in which a hydrogen atom is expected to form three bonds with the surface atoms should be, in principle taken into account. The sites of interest are labelled in Fig.1 as A, B, C and D.

Let us briefly focus on the first four sites. First of all, if a hydrogen atom occupies the site marked A which is situated above the tetrahedral void, one of three substrate atoms which take part in H-Pd bonds formation is positioned at the edge. As a result only two of three H-Pd bonds are saturated. Then, in the case of B-site two of three palladium atoms involved in H-Pd bonds formation are located at the edge. That is why the number of H-Pd saturated bonds is limited to one. Also, the site C is, in our opinion, very similar to those on Pd (111) for which the adsorption energy per hydrogen atom is equal to 0.51 eV [23]. It is clearly seen that each of the bonds which hydrogen atom being adsorbed in this position forms with substrate atoms should be saturated. Finally, if the hydrogen atom is chemisorbed on site D which is built into the terraces steps, it is located above the octahedral void and, like the case of B position, it can form two unsaturated bonds with the substrate atoms. However, the palladium atom which forms saturated bonds with a hydrogen atom adsorbed on B, is positioned just on the surface, whereas, in the case of D-site, the substrate atom which participates in the creation of an H-Pd saturated bond offers a lesser number of unsaturated bonds. Hence, we suppose that the B-site is the most favourable one for a hydrogen atom to be adsorbed. Then, the energetic ordering of the adsorption sites should be as follows: A, D and C.

The question arises, in what way we should assign the peaks observed in TED spectrum of electrons field emitted from the hydrogen covered Pd (133) surface (Fig.4b) to the available adsorption sites available on this surface. A partial answer can be given through the analysis of the enhancement factor R which, in the case of chemisorption studies, is given by the ratio of the TED $j_a^{'}(\varepsilon)$ from an adsorbate covered surface to the TED from the clean surface $j_o^{'}(\varepsilon)$ [24]:

$$R \equiv \frac{j_a^{'}(\varepsilon)}{j_0^{'}(\varepsilon)}.$$

According to the chemisorption theory [25], if an adsorbate atom approaches a metal surface, its sharp level broadens by interaction with the metal, and the enhancement factor can reveal the characteristics of the local density of states of the adsorbate.

Fig.5 shows the enhancement factor R for emission from Pd (133) surface exposed to a 240 L dose of hydrogen at $T = 78$ K.

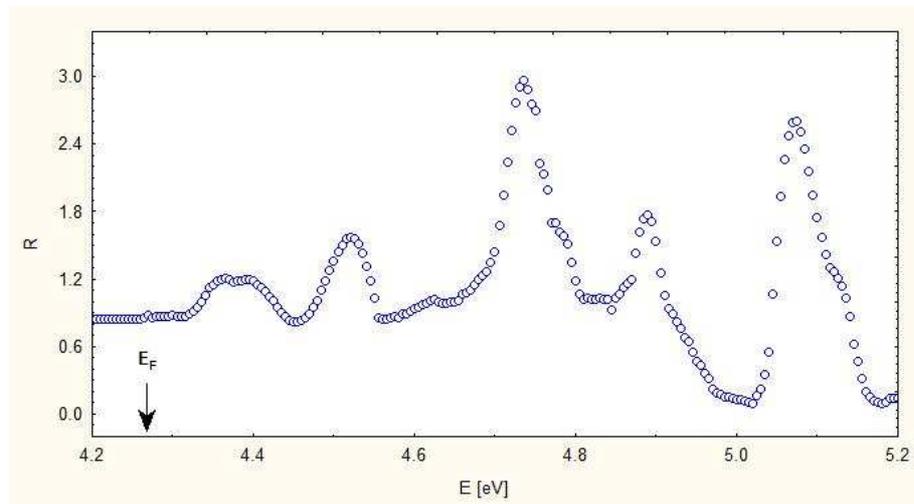

Fig.5 Electron enhancement factor R for hydrogen adsorption on Pd (133). The applied dose is 240 L.

It is seen from the figure that under these conditions a well-defined structure in R develops, which consists of a hump at about 4.4 eV as well as four peaks positioned at the energy 4.52; 4.74; 4.89 and 5.08 eV, respectively. In our opinion, the origin of the hump is completely different than the other peaks. We believe that the occurrence of the hump may be connected with the broadening of the Fermi level due to adsorbate-induced changes in electronic structure of the surface whereas the four peaks result from the broadening and shift of adatom's energy levels. Our assumption seems to be reasonable if we keep in mind the existence of four potential adsorption sites on the Pd (133). The full width at half-maximum (FWHM) of each peak as well as their intensities are collected in Table1.

Table1 Summary of the electron enhancement factor of the field electron emission from the hydrogen covered Pd (133) surface

| Number of peak | Energy (eV) | Intensity (a.u) | FWHM (eV) |
|---|---|---|---|
| 1 | 4.52 | 1.62 | 0.05 |
| 2 | 4.74 | 3.12 | 0.08 |
| 3 | 4.89 | 1.83 | 0.05 |
| 4 | 5.08 | 2.57 | 0.05 |

The above-mentioned effects, namely the broadening and shift of adatom's level are mainly due to the binding hydrogen-substrate interaction leading to the hydrogen adsorption on a specific site. However, the applied electric field is another factor which cannot be omitted in the appearance of the above-mentioned effects.

The significance of the presence of the field is even more pronounced if the adsorption on a stepped surface is taken into account. It is a well-known fact that the magnitude of a local electric field at the most protruding atoms positioned at the edges is enhanced in comparison with more smooth surface regions. Hence, because of the field presence, we should expect that hydrogen should be bound stronger with the sites labelled by A, B and D (Fig.1) than those with C. As to the influence of electric field on the shift of an adatom's energy levels, we believe that the higher electric field present during the interaction is the deeper the level is positioned below $E_F$. Therefore we shall assign the peak at 5.08 eV (Fig.5) to the adsorption of hydrogen on the site denoted by B on the Pd (133) surface. Then, the adsorption on the A-site results in the peak appearance at 4.89, while the peak at 4.74 is caused by the adsorption on the D-site. Finally, the peak at 4.52 eV is related to hydrogen adsorption on the C-site, which is considered as the least energetically favourable one among the other ones.

After a dose of 600 L was applied, the significant alterations in TED picture were noticed (Fig.4c). First of all, the appearance of two distinct peaks at 3.85 and 4.03 eV is remarkable. Also, in comparison with the former case, the shift of $E_F$ position, relating to the decrease of work function is observed. In our opinion, the above-mentioned changes are caused by the formation of surface palladium-hydrides structures.

This assumption is based on the results obtained by Duś and his co-workers [26], which contributed widely to our knowledge on the thermodynamics of transition-metal hydrides formation. On the basis of their simultaneously carried out measurements of the surface potential and hydrogen pressure, they have been able to give the following description of charge transfer within the adsorbate that accompanies the process of palladium hydrides formation at the temperature 78 K. Accordingly, at the beginning of adsorption the negatively charged adspecies referred to the $β^-$ state appear on the surface, which leads to the observed increase of work function. However, as the hydrogen coverage increases, the formation of positively charged adspecies (termed $β^+$) takes place. The nature of these adspecies is very similar to that of the ones in the bulk hydride taking part in the construction of the new electronic structure of $PdH_x$, which takes place. Hence, it is essential that $β^+$ adspecies incorporate easily into the subsurface and subsequently into the bulk of the substrate. Nevertheless, part of $β^+$ adspecies still remains on the surface, leading to the decrease of work function, which is also observed in our experiment. It is worth mentioning that the process of hydrides formation is accompanied by the rapid changes in FEM image (Fig.2d).

### IV. Conclusions

We have presented the results of TED of electrons emitted from the hydrogen covered Pd (133) surface. Depending on the applied hydrogen dose we have found that:
a) At 240 L hydrogen exposure the adsorption takes place on four different sites. This results in the appearance of four resonant states positioned at 0.32; 0.55; 0.70 and 0.88 eV below $E_F$.
b) At 600 L the shift of $E_F$ is observed as a result of surface palladium hydrides formation.

We have also identified the preferential adsorption sites on the Pd(133) on the basis of both the geometry considerations and the measurements of the enhancement factor. However, we would like to underline that calculations based on quantum chemistry should be done to verify our indication.

## V. Acknowledgement

The authors acknowledge the financial support of the Ministry of Science and Higher Education (Grant No. N207 017 31/0864).